\documentclass[editedvolume,numreferences]{crckbked}

\usepackage{graphicx}

\def \be{\begin{equation}}
\def \ee{\end{equation}}
\def \bmlett{\begin{mathletters}}
\def \emlett{\end{mathletters}}

\def \HH{{\mathcal H}}
\def \NN{{\mathcal N}}

\def \ra{\rightarrow}

\begin{document}
\begin{opening}
\title{Resonant Cooper-Pair Tunneling: Counting Statistics and Frequency-Dependent Current Noise}
\author{A.A. Clerk}
\institute{Departments of Applied Physics and Physics\\Yale
University, PO Box 208284, New Haven, CT 06520-8284}
\end{opening}

\section*{Abstract}
We discuss the counting statistics and current noise associated
with the {\it double} Josephson quasiparticle resonance point in a
superconducting single electron transistor.  The counting
statistics are in general phase-dependent, despite the fact that
the average current has no dependence on phase.  Focusing on
parameter regimes where the counting statistics have no
phase-dependence, we use a general relation first derived by
MacDonald in 1948 to obtain the full frequency-dependent shot
noise {\it directly} from the counting statistics, without any
further approximations.  We comment on problems posed by the
phase-dependence of the counting statistics for the
finite-frequency noise.

\section{Introduction}

Resonant Cooper-pair tunneling, also known as Josephson
quasiparticle tunneling, refers to transport cycles in
superconducting single-electron transistors (SSET's) which involve
the transfer of both Cooper pairs and quasiparticles
\cite{Fulton,Averin,Maasen}. They have recently been the subject
of renewed attention, both because of their unusual noise
properties \cite{Choi,Aash,Johansson} and because of their utility
in measuring the state of a charge superconducting qubit
\cite{Nakamura,Lehnert}. In terms of noise properties, it has been
shown that charge fluctuations associated with these processes can
induce a population inversion in a coupled two-level system (i.e
in terms of its charge noise, the transistor effectively has a
negative temperature) \cite{Aash,Johansson}. The shot-noise in the
current through the transistor was also found to have remarkable
properties \cite{Choi, Aash}. By tuning the strength of the Cooper
pair tunneling relative to the quasiparticle tunneling, one could
effectively tune the Fano factor determining the zero-frequency
shot noise. It was possible to achieve a Fano factor greater than
one, which was interpreted as a consequence of the effective
charge associated with the transport cycle being greater than one.
Perhaps more surprisingly, it was possible to reduce the Fano
factor below $1/2$, behaviour that was not fully explained.  The
finite frequency current noise also showed interesting behaviour
\cite{Choi}-- in the regime where the Cooper-pair tunneling
dominated the quasiparticle tunneling, a coherent peak in the
current noise was predicted at the Josephson energy.

In this paper, we look specifically at the statistics of
transferred charge and current noise associated with the {\it
double} Josephson quasiparticle (DJQP) process in a SSET, a
process which allows for near quantum-limited measurement
\cite{Aash}, and which has been used in a recent experiment
\cite{Lehnert}.  We pay careful consideration to the fact that the
breaking of gauge invariance by superconductivity can make the
interpretation of counting statistics more subtle than in the
normal-metal case \cite{Markus,Belzig,Shelankov}.  One obtains
phase-dependent counting statistics here despite the fact that
transport is insensitive to the overall phase difference between
the two superconducting reservoirs.  We identify two regimes in
which there is no ambiguity in defining counting statistics, and
proceed to use them to {\it directly} calculate the
frequency-dependent shot noise, without any further
approximations.  This is in contrast to previous treatments both
of normal and superconducting single electron transistors
\cite{KorotkovNoise,Makhlin,Choi}, where an additional Markov
approximation (beyond what is needed for the counting statistics)
is made to obtain the frequency-dependent current noise.  The
direct connection between counting statistics and
frequency-dependent current noise we employ here allows one to
give a simple interpretation of the latter.  In particular,
coherent peaks in the shot noise appearing at the Josephson energy
can be directly tied to the time dependence of the counting
statistics.  We also suggest that the suppression of the Fano
factor below $1/2$ is related to the coherence of Cooper-pair
tunneling.

\section{Description and Model for Josephson Quasiparticle Tunneling}

\begin{figure}
\centerline{\includegraphics[width=8cm]{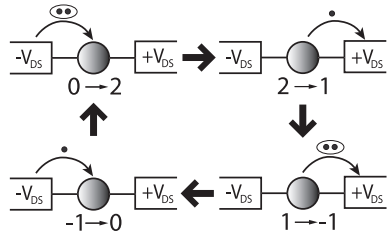}}
\caption{Schematic showing the four steps of the double Josephson
quasiparticle process which can occur in a superconducting
single-electron transistor. Circles represent the central island
of the SSET, while the rectangles are the electrodes.  Numbers
indicate the charge of the SSET island.} \label{EyeSchem}
\end{figure}

A SSET consists of a superconducting, Coulomb-blockaded island
which is attached via tunnel junctions to two superconducting
electrodes (Fig.~\ref{EyeSchem}). The SSET Hamiltonian $\HH_{S} =
H_{K} + H_{C} + H_{V} + H_{T}$ has a term $H_K$ describing the
kinetic energy of source, drain and central island electrons (i.e.
three bulk BCS Hamiltonians),  a term $H_V$ which describes the
work done by the voltage sources, and a tunneling term $H_T$. The
charging interaction is $H_C = E_{C} (n - \NN)^2$, where $E_{C}$
is the SSET capacitive charging energy, $n$ is the number of
electrons on the central island, and $\NN$ is the dimensionless
gate voltage applied to the island. We consider a SSET with
identical tunnel junctions, whose dimensionless conductance $g$
satisfies $g/(2 \pi) \ll 1$.

The DJQP process occurs when both the transistor gate voltage
$\NN$ and drain-source voltage $2 V_{DS}$ are tuned such that two
Cooper-pair tunneling transitions, one in each junction, are
resonant.  We label these transitions as $n = 0 \ra 2$ in the left
junction, and $n = 1 \ra -1$ in the right junction (see
Fig.~\ref{EyeSchem}).  The double resonance point for these
transitions occurs at $eV_{DS} = E_{CS}$ and $\NN_S = 1/2$. In
addition, $E_{C} / \Delta$ (where $\Delta$ is the superconducting
gap of the SSET) must be chosen so that the quasiparticle
transitions linking the two Cooper pair resonances are
energetically allowed (i.e. $n = 2 \ra 1$ and $n = -1 \ra 0$),
whereas unwanted transitions which would interrupt the cycle (i.e.
$n = 0 \ra 1$) are not. We take $E_{C} = \Delta$ to satisfy these
conditions; this corresponds to the experiment of Ref.
\cite{Lehnert}.

Assuming that all the above conditions are met, and that $T \ll
E_C$, transport through the SSET will be dominated by the DJQP
process sketched in Fig.~\ref{EyeSchem}.  The standard theoretical
description of this process \cite{Averin} is obtained by looking
at the dynamics of the reduced density matrix $\rho$ describing
the charge $n$ of the central SSET island.  Here, only four charge
states are important; further, off-diagonal terms need only be
retained between states involved in Josephson tunneling.  Letting
$\rho_{(i,j)} = \langle n=i | \rho | n=j \rangle$, we can
represent the non-zero elements of this reduced density matrix as
an $8$ component vector:
\begin{equation}
\vec{\rho} \equiv \left(
    \begin{array}{cccccccc}
      \rho_{(-1,-1)} & \rho_{(0,0)} & \rho_{(1,1)} & \rho_{(2,2)} &
      \rho_{(-1,1)} & \rho_{(1,-1)} & \rho_{(0,2)} & \rho_{(2,0)} \
    \end{array}
\right)
\end{equation}
By starting with the von Neumann equation for the evolution of the
full density matrix, treating the tunneling Hamiltonian to lowest
order in perturbation theory, and then tracing out the fermion
degrees of freedom (see, e.g. Ref. \cite{Makhlin}), one obtains
the an evolution equation for $\rho$.  Taking both junctions to be
on resonance, we have:
\begin{equation}
    \frac{d}{dt} \vec{\rho(t)}  =  U^{\dagger} M U \vec{\rho(t)}
    \label{EvoEqn}
\end{equation}
\begin{equation}
    M = \left(
    \begin{array}{cccccccc}
      -\Gamma & 0 & 0 & 0 & -i \frac{E_J}{2} & i  \frac{E_J }{2} & 0 & 0 \\
      \Gamma & 0 & 0 & 0 & 0 & 0 & -i \frac{E_J}{2} & i \frac{E_J}{2} \\
      0 & 0 & 0 & \Gamma & i \frac{E_J}{2} & -i \frac{E_J}{2} & 0 & 0 \\
      0 & 0 & 0 & -\Gamma & 0 & 0 & i \frac{E_J}{2} & -i \frac{E_J}{2} \\
      -i \frac{E_J}{2} & 0 & i \frac{E_J}{2} & 0 & -\Gamma/2 & 0 & 0 & 0 \\
      i \frac{E_J}{2} & 0 & -i \frac{E_J}{2} & 0 & 0 & -\Gamma/2 & 0 & 0 \\
      0 & -i \frac{E_J}{2} & 0 & i \frac{E_J}{2} & 0 & 0 & -\Gamma/2 & 0 \\
      0 & i \frac{E_J}{2} & 0 & -i \frac{E_J}{2} & 0 & 0 & 0 & -\Gamma/2 \
    \end{array},
\right) \label{BigM}
\end{equation}
\begin{equation}
    U = diag \left(1,1,1,1,e^{-i \phi_R},e^{i \phi_R},e^{-i
    \phi_L},e^{i \phi_L} \right)
\end{equation}
$E_J  = g \Delta / 8$ is the standard Ambegaokar-Baratoff value
for the Josephson energy emerging from perturbation theory in the
tunneling Hamiltonian; $\phi_{L}$ and $\phi_{R}$ are the phases of
the superconducting reservoirs. $\Gamma$ is the rate associated
with the two quasiparticle transitions occurring the DJQP cycle,
and is given by the usual expression for quasiparticle tunneling
between two superconductors \cite{Tinkham}; for simplicity, we
assume the two transitions to have equal rates.  We have also made
a Markov approximation to obtain Eq. (\ref{EvoEqn}), which is
valid as long we do not probe the evolution of $\rho$ on
timescales smaller than $\hbar / E_C$.

The zero eigenvector $\vec{\rho}_0$ of the evolution matrix $M$ in
Eq. (\ref{BigM}) corresponds to the stationary value of the
reduced density matrix describing the island charge; note that in
general, it will have non-zero off-diagonal elements.  The
diagonal elements of $\rho_0$ (which are probabilities) can be
obtained from a classical rate equation; one eliminates the
off-diagonal elements from Eq. (\ref{EvoEqn}) by expressing them
in terms of the diagonal elements. Letting $\vec{p}$ be the
$4$-vector of these probabilities, we are lead to the rate
equation:
\begin{equation}
    \frac{d}{dt} \vec{p}(t) =
        \left( \begin{array}{cccc}
          -\Gamma-\gamma & 0 & \gamma & 0 \\
          \Gamma & -\gamma & 0 & \gamma \\
          \gamma & 0 & -\gamma & \Gamma \\
          0 & \gamma & 0 & -\Gamma -\gamma \\
        \end{array} \right) \vec{p}(t) = 0
        \label{Incoherent}
\end{equation}
where $\gamma = E_J^2 / \Gamma$ represents a ``rate" for
Cooper-pair tunneling.  Eq. (\ref{Incoherent}) represents an
incoherent model for the DJQP process, and is sufficient for many
purposes (e.g., calculating the average current $\langle I \rangle
= 3 \gamma \Gamma / (4 \gamma+ 2 \Gamma)$). The full stationary
value of the density matrix (including off-diagonal terms) is
given by:
\begin{equation}
    \vec{\rho}_0 = \frac{1}{4 \gamma+ 2 \Gamma} \left(
        \gamma, \gamma+\Gamma, \gamma + \Gamma, \gamma,
        i E_J e^{-i \phi_R} ,-i E_J e^{i \phi_R} ,
        -i E_J e^{-i \phi_L} ,i E_J e^{i \phi_L}
        \right) \label{Rho0}
\end{equation}

\section{Counting Statistics}

In order to obtain information about the number of electrons that
have tunneled through one of the SSET junctions, the density
matrix approach of the previous section must be embellished.  The
standard approach for single electron transistors
\cite{Makhlin,Choi} is to employ a ``counter" scheme where one
explicitly tracks the dynamics of, say, $m_L$, the number of
electrons which have tunneled through the left junction of the
SSET. The Hilbert space of the SSET is expanded so that each state
is now also labelled by a value of $m_L$, and the tunnel
Hamiltonian is modified so that it raises or lowers this index as
appropriate (e.g., by introducing an auxiliary raising operator
$F^{\dagger}$ such that $[m_L,F^{\dagger}] = F^{\dagger}$). The
resulting equation for the reduced density matrix describing both
$n$ and $m_L$ is simplified if one Fourier transforms in the
latter variable. One finds that $\frac{d}{dt} \vec{\rho}(k) =
U^{\dagger} M(k) U \vec{\rho}(k)$, where we have defined:
\begin{eqnarray}
    \rho_{(n_1,n_2)}(k)  =
        \left\{
            \begin{array}{r@{\quad\quad}l}
                \sum_{m_L} e^{i k m_L} \langle n_1, m_L | \rho | n_2, m_L+2 \rangle
                    & \textrm{   if $n_1=0,n_2=2$,} \\
                \sum_{m_L} e^{i k m_L} \langle n_1, m_L+2 | \rho | n_2, m_L \rangle
                    & \textrm{   if $n_1=2,n_2=0$,} \\
                \sum_{m_L} e^{i k m_L} \langle n_1, m_L | \rho | n_2, m_L \rangle
                    &  \textrm{  otherwise.}
            \end{array} \right.
\end{eqnarray}
\begin{equation}
    M(k) = \left(
    \begin{array}{cccccccc}
      -\Gamma & 0 & 0 & 0 & -i \frac{E_J}{2} & i \frac{E_J}{2} & 0 & 0 \\
      \Gamma e^{i k} & 0 & 0 & 0 & 0 & 0 & -i \frac{E_J}{2} & i \frac{E_J}{2} \\
      0 & 0 & 0 & \Gamma & i \frac{E_J}{2} & -i \frac{E_J}{2} & 0 & 0 \\
      0 & 0 & 0 & -\Gamma & 0 & 0 & i \frac{E_J}{2} e^{2i k} & -i \frac{E_J}{2} e^{2 i k} \\
      -i \frac{E_J}{2} & 0 & i \frac{E_J}{2} & 0 & -\Gamma/2 & 0 & 0 & 0 \\
      i \frac{E_J}{2} & 0 & -i \frac{E_J}{2} & 0 & 0 & -\Gamma/2 & 0 & 0 \\
      0 & -i \frac{E_J}{2} & 0 & i \frac{E_J}{2} e^{-2 i k} & 0 & 0 & -\Gamma/2 & 0 \\
      0 & i \frac{E_J}{2} & 0 & -i \frac{E_J}{2} e^{-2 i k} & 0 & 0 & 0 & -\Gamma/2 \
    \end{array}
\right)
    \label{BigKM}
\end{equation}
The probability distribution for $m_L$ is thus easily obtained
from the characteristic function $p(k,t)$:
\begin{equation}
    p(m_L,t) = \int_{-\pi}^{\pi} \frac{dk}{2 \pi} e^{-i k m_L} p(k,t) \equiv
        \int_{-\pi}^{\pi} \frac{dk}{2 \pi} e^{-i k m_L} \sum_n \rho_{(n,n)}(k,t)
\end{equation}
\begin{equation}
    \vec{\rho}(k,t) = e^{M(k) t} \vec{\rho}(k,t=0)
    \label{RhoKT}
\end{equation}
It only remains to specify the initial state of the transistor
(i.e. $\vec{\rho}(k,t=0)$). The natural choice is to place the
transistor $n$ degree of freedom in its stationary state, and
choose $m_L=0$ with probability one (i.e. $\vec{\rho}(k,t=0) =
\vec{\rho}_0$) \cite{Makhlin}. $p(m_L,t)$ would then represent the
stationary probability that $m_L$ electrons are transferred
through the left junction in a time $t$. While this procedure is
perfectly well defined for non-superconducting systems
\cite{Makhlin}, it is problematic in the superconducting case.
The stationary state of the SSET has coherence between different
charge states-- thus, one cannot have both that $m_L=0$ {\it and}
the coherence required in the stationary state.  A possible remedy
for this problem would be to suppress the off-diagonal elements of
the initial density matrix; of course, this procedure needs to be
justified.  Note that given the linearity of Eq. (\ref{RhoKT}) in
the initial density matrix, we can separate incoherent and
coherent contributions to $p(m_L,t)$ (i.e. those arising,
respectively, from the diagonal and off-diagonal elements of the
initial density matrix).

To gain further insight, it is instructive to consider an
alternate to the ``counter" scheme discussed above. One can also
obtain a tunnelled charge distribution for our system by following
the general prescription for obtaining full counting statistics
discussed in Ref. \cite{Levitov,Markus}; we denote this
distribution as $p'(m_L,t)$. The procedure here involves
considering the effects of coupling an ideal measurer of current
to the transistor.  In practice, a gauge transformation is made to
eliminate the interaction with the measurement device, resulting
in a phase $e^{i \lambda(t)}$ being attached to the tunnel matrix
element $t$ \cite{Levitov2}.  Using the fact that the perturbative
calculation of $\rho$ in the tunneling Hamiltonian has a Keldysh
structure (see e.g. Ref. \cite{Makhlin}), one finds that
$\lambda(t) = \delta + k$ on the forward Keldysh contour, and
$\delta-k$ on the backwards Keldysh contour. $\delta$ is a phase
which depends on the initial state of the detector; for
non-superconducting systems, gauge invariance ensures that it does
not play a role. This procedure results in an expression for
$p'(k,t)$ which is identical in form to Eq. (\ref{RhoKT}), except
that now the initial state is uniquely specified as
$\vec{\rho}_0$, and the evolution matrix $M$ undergoes a
$\delta$-dependent unitary transformation which only affects the
coherent contribution to $p(k,t)$:
\begin{equation}
    M \rightarrow V^{\dagger} M V
    \hspace{1 cm}
    V = \textrm{diag} \left(
        1,1,1,1,1,1,e^{-i (k+\delta)},e^{-i (k-\delta)} \right)
\end{equation}
Thus, one has in general phase-dependent counting statistics for
our system, meaning that there is a dependence on the phase shift
$\delta$ introduced by the measurement.  This phase-dependence
exists despite the fact that the average current is not sensitive
to the phase difference between the two superconducting
reservoirs. As discussed in Ref. \cite{Markus,Belzig}, the results
of any charge-counting experiment will now be dependent on the
initial state of the detector.  We thus see that the ambiguity in
defining the initial state of the transistor in the ``counter"
scheme translates here to needing to know the initial state of the
detector.  Note that the incoherent part of $p(k,t)$ is the same
in the two schemes, while the coherent part differs.  The process
of ignoring the initial coherence of the density matrix in the
``counter" scheme corresponds here to assuming an initial detector
state which is completely delocalized in the phase $\delta$.

Despite the above caveats, we can determine the counting
statistics unambiguously in two limits where the off-diagonal
elements of the stationary state $\vec{\rho}_0$ are suppressed--
either $\Gamma \gg E_J$ or $E_J \gg \Gamma$ (c.f. Eq.
(\ref{Rho0})). In these limits the coherent contribution to
$p(k,t)$ vanishes, implying that the counting statistics are
identical in both schemes, and have no dependence on the phase
shift $\delta$. Note that in general, the magnitude of the
off-diagonal elements of $\rho$ in the stationary state are driven
both by the size of $E_J / \Gamma$ and by the population asymmetry
of the two charge states involved (i.e. $\rho_{02} \propto
(\rho_{00} - \rho_{22})$).  In the limit of large $E_J$, it is the
lack of population asymmetry which suppresses the off-diagonal
elements. In this limit, the symmetric and anti-symmetric
superpositions of charge states are equally populated, leading to
a vanishing of the off-diagonal matrix elements.

\subsection{The Limit $E_J \ll \Gamma$}

In this limit, the SSET effectively gets stuck in the states $n=0$
and $n=1$ waiting for the relatively slow Cooper-pair transitions
to occur; we can think of the rate $\gamma$ (c.f. Eq.
(\ref{Incoherent})) as describing effective transitions between
$n=0$ and $n=1$ (i.e. $\gamma$ describes both a Cooper-pair event
and the subsequent quasiparticle event). We find for the
probability distribution $p(m,t) \equiv p(m_L,t)=p(m_R,t)$ (i.e.
the junctions are identical) :
\begin{eqnarray}
    p(m,t) = \left\{
        \begin{array}{r@{\quad\quad}l}
            e^{-\gamma t} \frac{(\gamma t)^{2l}}{(2l)!}
                & \textrm {if $m=3l$}  \\
            \frac{e^{-\gamma t}}{2} \frac{(\gamma t)^{2 l+1}}{(2l+1)!}
                & \textrm{if $m=3l+1$ or $m=3l+2$}  \\
            0   & \textrm{otherwise.}
        \end{array} \right.
        \label{SmallEJDist}
\end{eqnarray}
We have neglected terms which are small as $(E_J / \Gamma)^2$, and
chosen $m=0$ at $t=0$.  Again, as the coherent contribution to
$p(m,t)$ vanishes to leading order, there is no ambiguity in
defining the counting statistics. The statistics can be given a
simple interpretation. Each Cooper-pair plus quasiparticle event
is described by the rate $\gamma$ and a Poisson distribution.
After an even number of $\gamma$ transitions, a multiple of 3
electrons must have been transferred through a junction. However,
after an odd number of $\gamma$ transitions, there is an equal
probability of having had an extra two electrons transferred (i.e.
if we are looking at the left junction, $n=0 \ra 2 \ra 1$) or of
having a single extra electron transferred (i.e. $n=1 \ra -1 \ra
0$ for the left junction). Note that there is an asymmetry here
which favours values of $m$ which are an integer multiple of $3$;
this is the analog of the even-odd asymmetry found for the single
JQP process \cite{Choi}.
Also, note that in this limit the counting statistics are
identical if we use the incoherent model of Cooper-pair tunneling
described by Eq. (\ref{Incoherent}); coherence plays no role.

\subsection{The Limit $E_J \gg \Gamma$}
In this limit,  we can think of there being long periods of
coherent Josephson oscillations which are interrupted by
infrequent quasiparticle transitions; note that we still assume
$E_J \ll E_C = \Delta$, so that it does not effect the energetics
of quasiparticle tunneling. Defining the Poisson distribution
$\Lambda(l)$ as:
\begin{equation}
    \Lambda(l) = e^{-\Gamma t/2} \frac{ (\Gamma t/2)^l}{l!}
\end{equation}
we find up to terms of order $E_J / \Gamma$:
\begin{eqnarray}
    p(m,t) = \left\{
        \begin{array}{c@{\quad\quad}l}
            \frac{1}{4}\left(
                \Lambda(2l)
                \left(3 + \frac{ \cos (E_J t) }{2^{2l}} \right)
                + \Lambda(2l+1) + \Lambda(2l-1)
            \right)
                & \textrm{ if $m = 3l$,}  \\
            \frac{1}{4} \left(
                \Lambda(2l+1) + \frac{1}{2} \Lambda(2l+2) \left(
                1 -  \frac{ \cos (E_J t) }{2^{2l+2}} \right)
            \right)
                & \textrm{ if $m = 3l+1$,} \\
            \frac{1}{4} \left(
                \Lambda(2l+1) + \frac{1}{2} \Lambda(2l) \left(
                1 -  \frac{ \cos (E_J t) }{2^{2l}} \right)
            \right)
                & \textrm{ if $m = 3l+2$,}  \\
            0
                & \textrm{ otherwise}
        \end{array} \right.
    \label{LargeEJDist}
\end{eqnarray}
where again, we have assumed an initial state with $m=0$ without
any ambiguity, as the coherent contribution to $p(m,t)$ is
negligible.  We can again think of the quasiparticle transitions
as effectively being described by a rate $\gamma = \Gamma/2$ and a
Poisson distribution.  Now, however, after an even number of
transitions (say $2 l$), it is not certain that $3 l$ electrons
will have been transferred through a junction; rather, due to
weakly damped Josephson oscillations, there is also a probability
to find $3l \pm 2$ electrons transferred.  Even more remarkably,
there are signatures of coherent oscillations in $p(m,t)$, this
despite the fact that the initial coherence of the density matrix
is irrelevant.  As we will see in the next section, the
time-dependence of the counting statistics has a direct impact on
the frequency dependence of the current noise. Note that the
oscillation serves to modulate the asymmetry which favours $m$
being a multiple of three.  For $\Gamma t \gg 1$, the probability
of $m$ being a multiple of $3$ is given by:
\begin{equation}
    p(m=3l) \ra \frac{5 + \cos(E_J t) }{8}
\end{equation}

\section{Current Noise and Counting Statistics}

In this section, we calculate the frequency dependent current
noise $S_I(\omega)$ associated with the DJQP process.  As is
standard, the capacitance of the junctions $C_L$ and $C_R$ may be
used to connect this quantity (which includes the effects of
displacement currents) to the noise associated with the tunneling
currents in each contact \cite{KorotkovNoise}:
\begin{equation}
    S_{I}(\omega) = \frac{C_L^2}{C_\Sigma} S_{LL}(\omega) +
        \frac{C_R^2}{C_\Sigma} S_{RR}(\omega) +
        \frac{C_L C_R}{C_\Sigma} \left(
            S_{LR}(\omega) + S_{RL}(\omega) \right)
\end{equation}
where
\begin{equation}
    S_{\alpha \beta} =  \lim_{T \ra \infty} \lim_{t_0 \ra \infty}
        \int_{-T}^{T} dt e^{i \omega t}
        \langle \langle \{ I_{\alpha}(t+t_0), I_{\beta}(t_0) \} \rangle
        \rangle.
\end{equation}
The order of limits here ensures that we are taking averages with
respect to the stationary density matrix of the SSET; note also
that we are calculating the classical part of the current noise,
which is a symmetric function of frequency. Previous calculations
of this quantity for both normal and superconducting transistors
have effectively made use of the so-called quantum regression
theorem, which involves making an additional Markov approximation
beyond that necessary to calculate the reduced density matrix
$\rho$ \cite{KorotkovNoise,Makhlin,Choi}.  It turns out that this
is not necessary; one can directly connect the particle-current
noise correlators to the counting statistics using a formula first
derived by MacDonald \cite{MacDonald} . Using the definition
$I_\alpha(t) = \partial_t m_{\alpha}(t)$, one can show:
\begin{equation}
    S_{\alpha \beta}(\omega) = 2 e^2 \omega \int_0^{\infty} dt
        \left( \sin \omega t\right)
        \frac{d}{dt} \langle \langle m_{\alpha}(t) m_{\beta}(t) \rangle
        \rangle,
        \label{MacFormula}
\end{equation}
where at $t=0$, $m_L = m_R = 0$ with certainty, and the system is
described by the stationary density matrix $\rho_0$.  The integral
in this expression should be interpreted within the theory of
distributions, i.e. $\int_0^\infty dt \sin \omega t = 1/\omega$.
We see from Eq. (\ref{MacFormula}) that the frequency-dependent
current noise is directly tied to the time-dependence of the
second moment of the tunneled charge distribution.  At long times,
$\langle \langle m(t)^2 \rangle \rangle \ra f I t / e$, where $f$
is the Fano factor; this leads to the usual expression $S_{\alpha
\alpha}(\omega = 0) = 2 e f I$. A frequency-dependent $S(\omega)$
indicates that $\langle \langle m(t)^2 \rangle \rangle$ deviates
from this linear in time behaviour at short times.  We thus see
that in addition to being useful for calculations, Eq.
(\ref{MacFormula}) provides a straightforward way to interpret
frequency-dependent current noise in terms of counting statistics.
Note that unlike approaches using the quantum regression theorem
\cite{KorotkovNoise, Choi}, one does not need to add an ad-hoc
term to Eq. (\ref{MacFormula}) to obtain the correct noise in the
high-frequency limit.

\begin{figure}
\centerline{\includegraphics[width=10cm]{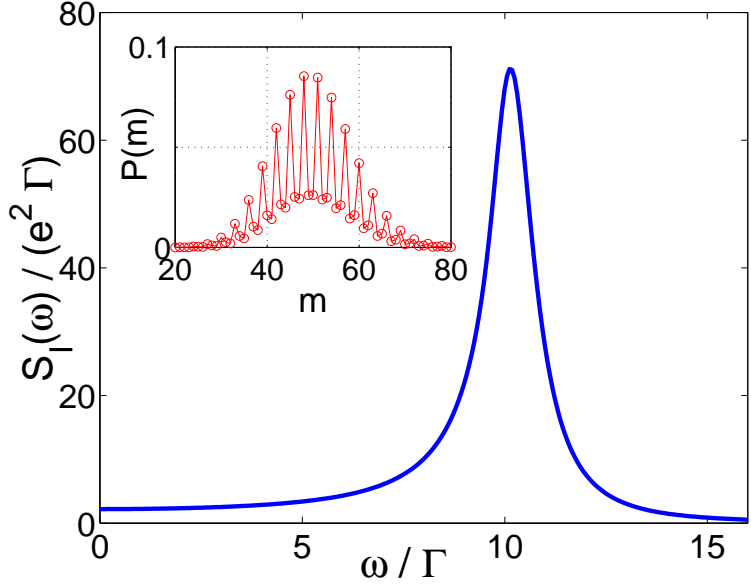}}
\caption{Frequency dependent current noise for the DJQP process at
$E_J = 10 \Gamma$; the large peak at $\omega = E_J$ is the result
of oscillations in the asymmetry of the tunneled charge
distribution. Inset: Distribution of tunneled charge $p(m)$ for
the same parameters at $\Gamma t = 67$; the asymmetry which
favours multiples of $3$ is clearly visible.} \label{LargeEJPlot}
\end{figure}

We apply Eq. (\ref{MacFormula}) to the DJQP process in the limit
$E_J \gg \Gamma$, where the counting statistics are well-defined
and phase-independent.  As noted in the previous section, the
coherent Josephson oscillations at frequency $E_J / \hbar$
modulates the asymmetry in $p(m,t)$ which favours multiples of
$3$.  This in turn modulates the second moment of the distribution
(i.e. the greater the asymmetry, the smaller the width); one finds
from Eq. (\ref{LargeEJDist}):
\begin{equation}
    \langle \langle m^2(t) \rangle \rangle =
        \frac{9 \Gamma t}{8}  + \frac{9 - e^{-\Gamma t}}{8}
        - \frac{\cos(E_J t)}{2} \left(
            e^{-\Gamma t/4} - e^{-3 \Gamma t / 4} \right)
\end{equation}
Eq. (\ref{MacFormula}) then implies that there should be a peak in
$S_{LL} = S_{RR}$ at $\omega = E_J$ of order $e^2 E_J^2 / \Gamma$,
which is much larger than the zero frequency noise $(9/4)e^2
\Gamma$. Shown in Fig. 2 are results obtained for $S_I(\omega)$
for $E_J = 10 \Gamma$, $C_L = C_R$; a sharp peak is indeed
visible. Similar results for the single JQP process were found in
Ref. \cite{Choi}.  We see here that the peak in the current noise
is directly related to the modulation in time of the asymmetry of
the counting statistics by coherent Josephson oscillations.

\section{The Intermediate Regime $E_J \simeq \Gamma$}

The noise properties in the regime $E_J \simeq \Gamma$ are
especially interesting, as it is in this regime where the DJQP
process can be used to make a near quantum-limited measurement
\cite{Aash}.  In this regime, the off-diagonal elements of the
stationary density matrix $\vec{\rho}_0$ are by no means small,
and thus the coherent, phase-dependent contribution to the
counting statistics will play a role.  While this in itself is not
a problem, the resulting status of $S_{I}(\omega)$ as calculated
using Eq. (\ref{MacFormula}) becomes unclear, as the quantity
$\partial_t \langle \langle m^2(t) \rangle \rangle$ is phase
dependent.  The conclusion would thus appear to be that for
resonant Cooper-pair tunneling, {\it the finite-frequency current
noise is phase dependent}, and consequently is sensitive to
details of the measurement.

We close by pointing out that the value of the zero frequency
noise remains phase independent and unambiguous regardless of the
ratio $E_J / \Gamma$.  At zero frequency, each of $S_{\alpha
\beta}$ are equal; thus, we only need to know $\partial_t \langle
\langle m^2(t) \rangle \rangle$ in the large time limit.  As
discussed in Ref. \cite{Makhlin}, this is determined completely by
the lifting of the zero eigenvalue of the evolution matrix
$M(k=0)$ in Eq. (\ref{BigM}) by non-zero $k$; the contribution to
$\langle \langle m^2(t) \rangle \rangle$ from other eigenmodes of
$M$ are damped away in the long time limit.  Restricting ourselves
to only this ``lifted" zero-mode, we have:
\begin{equation}
    p(k,t) \ra e^{\lambda_0(k) t} (1 + A(k,\delta)),
\end{equation}
where $A(k,\delta)$ contains all $k$-dependent terms involving the
initial density matrix, and the $\delta$-dependent eigenvectors
(left and right) corresponding to $\lambda_0(k)$; $A$ vanishes for
$k \ra 0$.  As:
\begin{equation}
    \langle \langle m^2(t) \rangle \rangle = - \frac{d^2}{dk^2}
    \log p(k,t)
\end{equation}
it follows from Eq. (\ref{MacFormula}) that
\begin{equation}
    S_{I}(0) \equiv S_{\alpha \alpha} = 2 e^2 \left( - \frac{d^2}{dk^2}
    \lambda_0(k) \Big|_{k=0} \right)
\end{equation}
The phase-dependent term $A$ is thus explicitly seen to play no
role. Analyzing the eigenvalue $\lambda_0(k)$ of $M(k)$ in Eq.
(\ref{BigKM}), one find for the Fano factor $f$ \cite{Aash}:
\begin{equation}
    f = \frac{3}{2} \left[
        1 - \frac{6 E_J^2 \Gamma^2}{\left( \Gamma^2 +
        2 E_J^2 \right)^2} \right]
\end{equation}
In both the limits $E_J \ll \Gamma$ or $E_J \gg \Gamma$ , $f \ra
3/2$.  This can be understood as an effective charge-- each
Cooper-pair plus quasiparticle transition transfers on average
$3/2$ electrons per junction.  Interestingly, when $E_J \sim
\Gamma$, $f$ drops below $1/2$, reaching a minimum of $3/8$ when
$E_J = \Gamma / \sqrt{2}$.  This behaviour is reminiscent of
double tunnel junction systems, or of a normal SET; in both cases,
there are two rates involved in transport, and the Fano factor
reaches a minimum of $1/2$ when these rates are equal. The
behaviour in such systems can be understood completely classically
by the self-averaging that occurs when one sequentially combines
two independent Poisson processes.  In contrast, the fact that $f$
drops below $1/2$ for the DJQP process is a direct consequence of
the coherence of Cooper-pair tunneling. To underscore this point,
one can calculate $f$ using the incoherent model of Eq.
(\ref{Incoherent}).  While this procedure yields the correct value
of $\langle I \rangle$ and the correct value of $f$ for extreme
values of $E_J / \Gamma$, it only gives a minimum $f$ of $1/2$.
One finds:
\begin{equation}
    f_{\textrm{incoherent}} = \frac{3}{2} \left[
        1 - \frac{2 E_J^2 \Gamma^2}{\left( \Gamma^2 +
        2 E_J^2 \right)^2} \right] =
        f +
        \frac{6 E_J^2 \Gamma^2}{\left( \Gamma^2 +
        2 E_J^2 \right)^2}
\end{equation}
Note that the incoherent calculation of $f$ is always an
overestimate; as expected, {\it coherence between charge states
suppresses the zero-frequency current noise}, as it tends to make
tunnelling events more regular.

\section{Conclusions}

We have studied the counting statistics and finite-frequency shot
noise of the DJQP process in a superconducting single electron
transistor, attempting to clarify some of the remarkable noise
features found in previous studies.  In general, the counting
statistics are phase-dependent, despite the fact that the average
current has no phase dependent.  An interesting question remains
how this phase dependence impacts the finite frequency noise,
given Eq. (\ref{MacFormula}) which directly relates the two
quantities.

\begin{acknowledgements}
I am grateful to S. M. Girvin for useful discussions, and to L.
Fedichkin for bringing Ref. \cite{MacDonald} and Eq.
\ref{MacFormula} to my attention. This work was supported by the
NSA and ARDA under ARO contracts ARO-43387-PH-QC, by the NSF under
DMR-0196503 \& DMR-0084501, and by the W.M. Keck Foundation.
\end{acknowledgements}

\end{document}